# Blockchain of Signature Material Combining Cryptographic Hash Function and DNA Steganography

Yixin Zhang

B CUBE, Center for Molecular Bioengineering, Technische Universität Dresden, Dresden, Germany

**Abstract**

An ideal signature material and method, which can be used to prove the authenticity of a physical item and against forgery, should be immune to the fast developments in digital and engineering technologies. Herein, the design of signature material combining cryptographic hash function and DNA steganography is proposed. The encrypting materials are used to construct a series of time-stamped records (blockchain) associated with published hash values, while each DNA-encrypted block is associated with a set of DNA keys. The decrypted DNA information, as digital keys, can be validated through a hash function to compare with the published hash values. The blocks can also be cross-referenced among different related signatures. While both digital cryptography and DNA steganography can have large key size, automated brutal force search is far more labor intensive and expensive for DNA steganography with wet lab experiments, as compared to its digital counterpart. Moreover, the time-stamped blockchain structure permits the incorporation of new cryptographic functions and DNA steganographies over time, thus can evolve over time without losing the continuous history line.

Because of the fast developments in digital and engineering technologies, many methods used to be considered as safe and secure have become increasingly questionable, including signature and stamp for self-identification. For 3D objects, additive manufacturing (3D printing)[1-2] represents a revolution in mechanical engineer, however, will also make counterfeit not only remarkably easier but also drastically faster and cheaper. Digital technology would not provide the ultimate solution, which it has helped to create. For example, the power and utility of cryptographic hash function[3] and prime factorization are based on the hypothesis that parallel computation will not be realized in the near future. Therefore, for long-term security challenges, making hash values with more bits would not provide the final solution. What's unbreakable today might be very different after hundreds and thousands of years. Although data safety over a long period of time is not relevant for most of our daily activities, it could be very critical for proving the authenticity of physical items, e.g. an important document, an art work, or an item of historical significance.

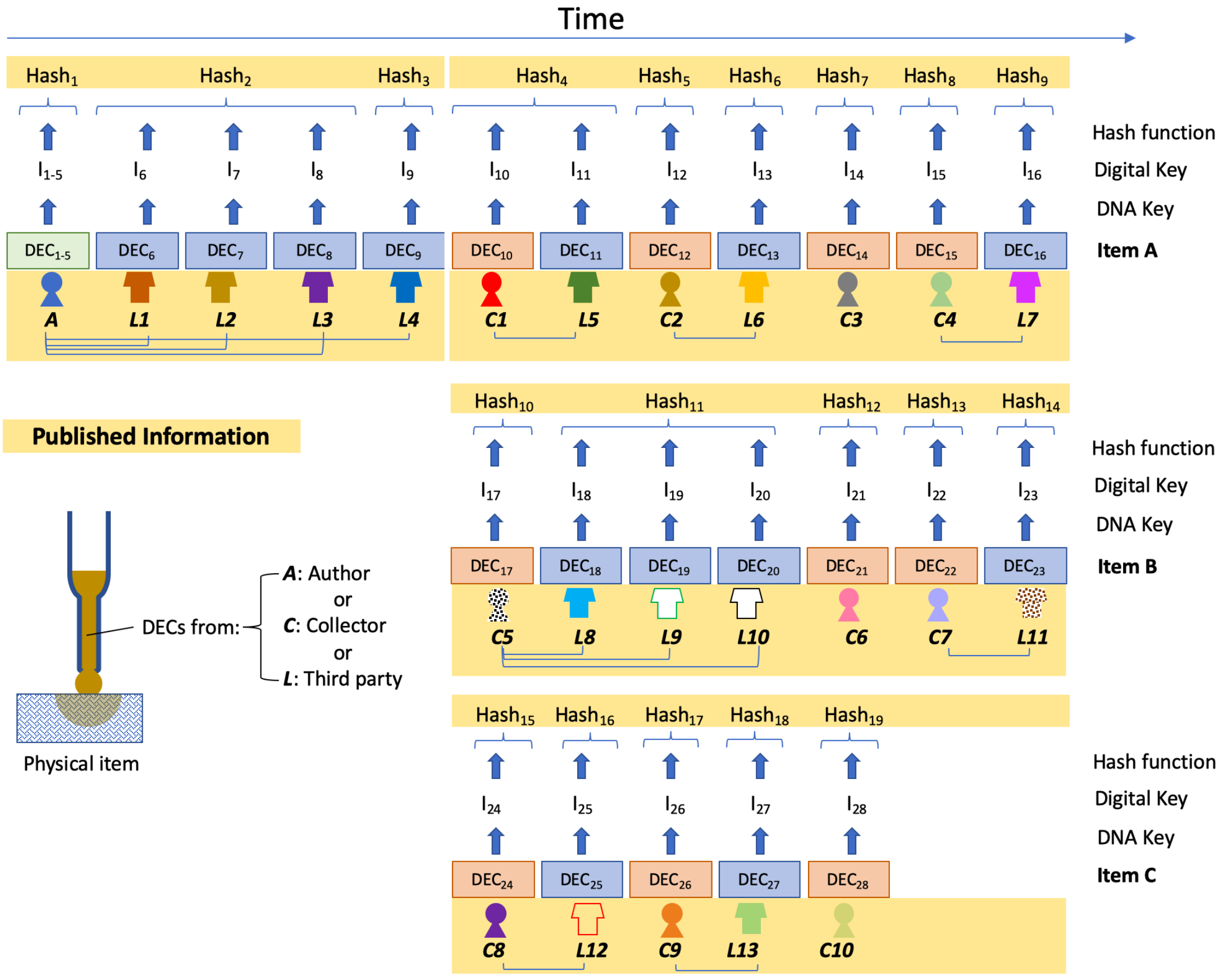

*Figure **1**. **Hybrid signature strategy.** DNA-based materials containing one or multiple DECs will be added to a physical item to be certificated. In this example, the $DEC_1$-$DEC_9$ are shared by all three items. The author as well as the collectors will generate their own DECs, as well as acquire DECs from third parties. The connection between a third party with the author or a collector will be published. Each DEC can be decrypted by a set of DNA keys, to reveal the DNA information, which will be used as digital key. The digital keys will be validated through a cryptographic hash function, while the hash values were previously published. Over time, new DECs will be added to the item by future collectors, to establish a blockchain. Examples of generating $Hash_1$, $Hash_2$ and $Hash_3$ are shown in the supporting information.*

Molecular signatures, which can be easily incorporated into various physical objects (e.g. ink, paper, texture, etc), are attractive for developing new certification tools. As novel data storage[4-10] and encryption medium[11-14], DNA possesses high capacity and longevity, and can be amplified and operated biochemically to perform computational tasks[15-18]. Herein, a signature material combining cryptographic hash function with DNA steganography (figure **1** and **2a**) is proposed. Encrypted DNA materials are added to the item to be certificated as molecular signature. To decrypt the DNA-encoded information needs to perform biochemical experiments with specific knowledge (DNA keys), and the revealed DNA information will be used as digital keys. The digital keys will be validated through a cryptographic hash function (e.g. SHA-256 or SHA-512), while the hash values were previously published (Figure **2a**). Therefore, to prove the authenticity of an item, people must be able: **i**) to extract signature materials from the item; **ii**) to use a set of DNA keys to generate digital keys experimentally; and **iii**) to validate the digital keys (Figure **2a**).

As a technology complimentary to digital cryptography to improve data safety, the DNA steganography can be as complex as compute-based algorithms such as SHA-256 and SHA-512. More importantly, decryption of DNA steganography requires to perform biochemical experiments with the physical item of interest in wet labs. This process will be much more labor intensive, thus limiting the use of automated brutal force search, which is drastically easier for silicon-base computation than for wet lab experiments. The DNA steganography method used in the paper possesses a QKD (Quantum Key Distribution)-like function and a self-destructive mechanism (Figure **2b**)[14]. QKD is based on the feature that the process of measuring a quantum system disturbs the system[19-20]. A solution of DNA molecules does not possess the property of quantum superposition. However, if an attempt of measurement can change its composition and leave a permanent trace, the design of DNA steganography can be considered as QKD-like. The QKD-like function and self-destructive feature can prevent un-authorized sample replication, thus prohibiting high throughput testing of the amplified DNA materials. Furthermore, the protocol includes a blockchain mechanism[21], allowing to improve information safety over time.

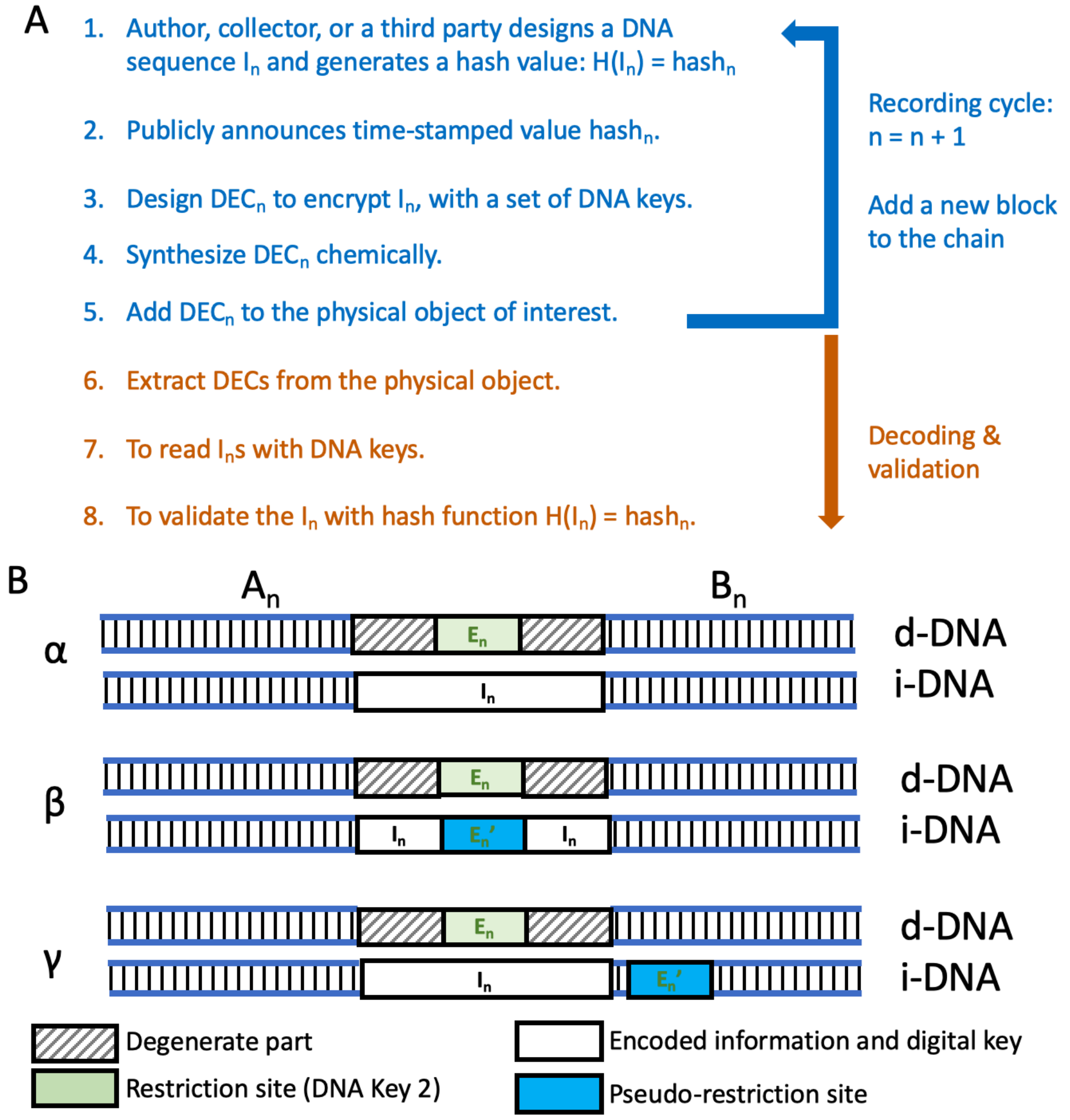

*Figure* 2. ***Encoding/decoding protocol and DEC design***. ***A****) DECs are designed, synthesized, and added to the item to be certificated over time, to establish a blockchain.* ***B****) Each DEC is a mixture of a d-DNA, and an i-DNA. To prevent to be analyzed by high throughput sequencing, DECs are pre-mixed with genomic DNA before applying to the item to be certificated. Each DEC can be decoded by applying the DNA key 1 (primers $A_n$ and $B_n$) and DNA key 2 (restriction enzyme $E_n$). The pseudo-restriction sites $E_n$'s are restriction sites different from $E_n$, to prevent digestions using a mixture of enzymes. When a wrong enzyme is used, the i-DNA will be destroyed.*

## DEC (DNA encrypting constructs) design

As a DNA-based signature material, a DEC contains the information about the DNA keys, which are used to reveal the encrypted digital key (DNA sequence as $I_n$ in figure **1**). A randomness-incorporating DNA steganography[14] was applied to design the DECs, which needs two types of DNA keys for decrypting each DEC, including a pair of DNA primers and a restriction enzyme (Figure **2b**). The message (the sequence of DNA $I_n$, which will later be used as the digital key) is flanked by two primer sequences ($A_n$ and $B_n$) and the resulting i-DNA is mixed with d-DNA (in large excess). The d-DNA possesses the same length and shares the same primers as i-DNA, thus can function as a mask to cover the i-DNA, even when the primers for a certain DEC are known. The middle region of d-DNA contains a restriction site $E_n$ flanked by two randomized sequences. For example, if the middle region of the d-DNA is 20 bp and the restriction site is 6 bp, the resulting d-DNA presents a $4^{14}$ sequence-diversity. The randomness is important for the QKD-like function and self-destruction feature, as discussed later.

The message $I_n$ can be revealed through sequencing after digesting the d-DNA with the right restriction enzyme and amplification with the right primers $A_n$ and $B_n$. Without the right restriction enzyme pre-treatment (DNA key-2), heating associated with PCR amplification will cause reshuffling among many different d-DNA sequences and i-DNA sequence. Consequently, the restriction site flanked by mismatches cannot be recognized by restriction enzyme. Eventually, this process will leave a permanent trace (QKD-like function) and make the information unreadable (self-destruction), even when the sample is afterwards subjected to the right processing.

This DNA key-2 of the basic DEC format $\alpha$ could be possibly bypassed when a mixture of restriction enzymes is applied, though different enzymes often need different specific buffer conditions. This can be circumvented by including a pseudo-restriction site $E_n$' in the i-DNA strands, either in the information part or in the primer sequence, as shown in the DEC formats $\beta$ and $\gamma$, respectively. To prepare the signature materials of DECs, the synthetic DNA sequences will be premixed with large excess of genomic DNA, which can cover the DECs and prevent them from analysis using next generation sequencing.

**Key sizes of DECs**

Every DEC has two types DNA keys: Key-1 is the two primers and Key-2 is the restriction enzyme. The optimal length for PCR primers is 18-22 nt. As a very conservative estimation, each prime with > 16 correct nt would be sufficient for PCR and the two primers are equivalent to $4^{16} \times 4^{16}$ = 64 bits of information. Additionally, people will need the right restriction enzyme for reading the information, which would add an additional > $2^7 - 2^8$ probability (>7 bit, for about 200 enzymes). As shown in figure **1**, the author's information is encrypted in (64+7) x 5 = 355 bits ($Dec_1$ to $DEC_5$). Not mentioning the laborious and expensive experiments with DNA, such information in their digital form cannot be decrypted through automated brutal force search. With the addition of new DECs to the item to grow the blockchain over time, the resulting system will become increasingly difficult for decryption.

**Encoding protocol**

The author will add his DECs and the DECs acquired from third parties (e.g. law firm or connoisseur) to the item to be certificated (e.g. DEC1-9 in figure **1**). A third party plays a role similar to block creators (for digital currencies such as bitcoin) and is rewarded for their activities to provide as well as to validate the DECs. To further grow the blockchain, the collectors will add their DECs and the DECs from third parties (e.g. DEC10-16 in figure **1**). The associations between author or collectors with their related third parties will be published. $Hash_1$ is generated with $I_1$-$I_5$ (e.g. through SHA-256 or SHA-512). The new DNA-encoded information, together with the hash value *n-1* (figure **3**), will be used to generate hash value *n*. While DNA keys and digital keys will be kept secret (by author, collector, and third party), the hash values will be published.

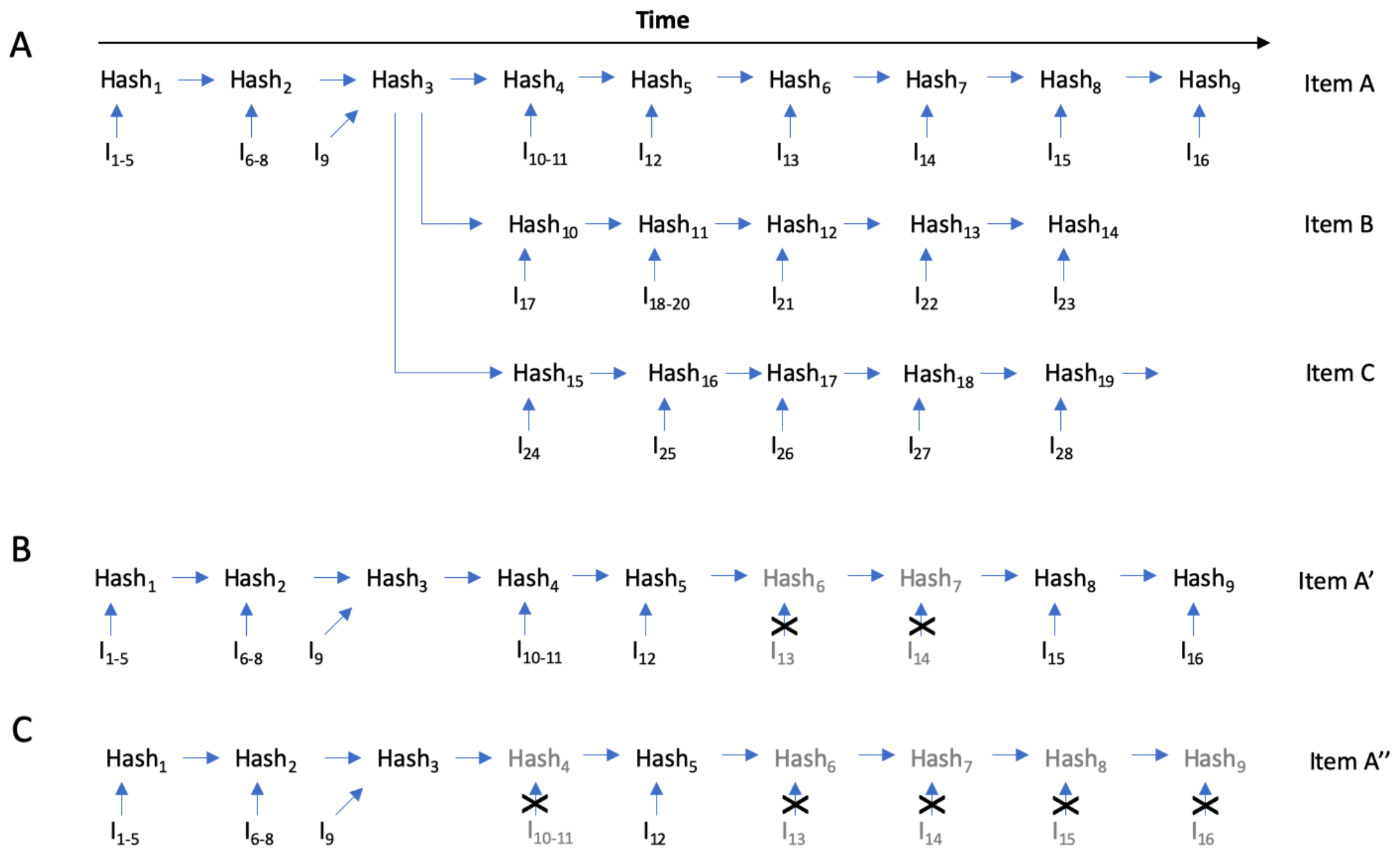

*Figure **3**. **Development of blockchain**. A) $Hash_1$ is generated with $I_1$-$I_5$. The new digital keys, together with the hash value n-1, will be used to generate hash value n. **B**-**C**) Some information will be lost over time. An item that can provide the most complete block chain, in full or the closest agreement with the published records, should be considered with the highest confidence as authentic. For example, the item A’ is more likely to be authentic than A”.*

The author will generate a set of DECs for his own, as well as acquiring a set of DECs from one or multiple third parties. The author will pass his DNA keys and digital keys, but not those from the third parties, to the collector #1. This will allow future collector (e.g. collector #2) to check the authenticity with the third parties. This is in the interest of the author, because this will prevent the collector #1 to make counterfeits based on the obtained information. The collector #1 will add his DEC, with the option of adding the DECs from one or multiple third parties. The collector #1 will pass the author’s and his keys to the collector #2. The collector #1 will not pass the keys from the third parties to collector #2. This is in the interest of the collector #1, because this will prevent the collector #2 to use collector #1’s credibility to make counterfeits in the future. A hash value can be generated either by a single digital key (e.g. $H(hash_4, I_{12}) = Hash_5$, where $H(x)$ is a hash function) or with multiple digital keys (e.g. $H(hash_3, I_{10}, I_{11}) = Hash_4$). The latter is certainly saver than the former.

## Discussion

In ancient China, people have developed an elegant method to prove the authenticity of art works, most commonly for paintings and calligraphies. The author put his stamp, often multiple stamps onto the piece. In the following years and centuries, both collectors and connoisseurs will add their stamps to the piece. Because their stamps can also be found in many other documents, they can be cross-referenced with each other. Therefore, the more cross-referencing stamps can be found on a piece, the higher confidence regarding the authenticity will a collector have. Similarly, a blockchain, which leads to the creation of bitcoin and many other digital currencies, is a growing list of records, called blocks, that are linked using cryptography. Each block contains a cryptographic hash of the previous block, a timestamp, and transaction data. The blockchain method is the digital counterpart of the stamp-chain, based on the assumptions that it is very difficult to fake multiple stamps (in ancient time), as well as to reverse password encryption of a one-way hash function (with modern computational power).

Using the protocol described in this paper, the authenticity can be confirmed with the presence of **P**hysical item (e.g. hand-written signature), the biochemical experiments with DNA keys to reveal a set of **D**NA sequences, as well as the use of the DNA sequences as digital keys to produce the **H**ash values (P-D-H protocol). The P-D-H protocol allows people to add new DECs, thus increasing security over time. The author can generate a set of DECs as a general signature for all of his works and documents, as well as a set of DECs for a particular item (figure **3a**). Two different items from the same author will lead to two different block chains, even if the author had signed them with the same set of DECs. Two different items will pass through different collectors, who will add their DECs (from themselves and/or from third parties). Some information will be lost over time: ink will fade, DNA can degrade, and the DNA and digital keys could also be lost when passing through many generations, as well as after closing law firms (as third party). However, the item that can provide the most complete block chains, in full or the closest agreement with the published records, should be considered with the highest confidence as authentic. For example, the item A' (figure **3B**) is more likely to be authentic than A'' (figure **3C**) as the block chain is relatively more complete for A'. Moreover, the information of early digital and DNA keys ($DEC_{1\text{-}9} \rightarrow I_{1\text{-}9}$) are associated with many works from the author, thus have the highest probability to be leaked, allowing forger to fake the signature materials.

DNA sequencing is more error-prone than copying digitally. Nevertheless, very little computational effort would be required to correct a digital key, while the brutal force search for the entire digital key remains impossible. For example, to correct < 3 mistakes in a 100 nt DNA sequence (corresponding to 2 or 4 bits in 200 bits of information) needs less than $300 + 9 \times {}_{100}C_2 = 44850$ times of computation.

The DECs have been designed to achieve high safety. For example, the combination of $DEC_1$ to $DEC_5$ generates a 355-bit DNA keys, while the third parties 1-4 produce a 213-bit ($Hash_2$) and a 71-bit ($Hash_3$) DNA keys (figure **1**). In the following scenario, their security can be partially compromised by:

- **Collectors.** Collector #1 can fake the author's DEC, but not those from third parties, as well as those from future collectors. For example, collector #1 does not have the 213-bit DNA key ($Hash_2$) and 71-bit DNA key ($Hash_3$).

**- *de novo* decryption.** A person (X) who gets the signature materials can analyze it and try to guess the DNA keys. X will have the opportunity to test all possible primers and restriction enzymes (DNA keys) as a whole to find the digital keys. The security is compromised, for example, reducing the 355-bit DNA key associated with $hash_1$ to 5 independent 71-bit DNA keys, though it remains experimentally impossible to perform brutal force search. The randomness-incorporating DNA steganography method used in this paper possesses a QKD-like function and self-destruction feature, thus preventing the amplification of signature materials to perform high throughput testing (brutal force search). Moreover, because the P-D-H protocol possesses a mechanism to increase security over time, newer and safer DNA steganography developed in the future can be incorporated into the blockchain.

- **Reverse decryption.** A person (X) can try to obtain the digital keys through computational brutal force search. However, given that different inputs can lead to the same output with a hash function, the identified digital key $I_x$ can be different from $I_n$. Even if $I_x$ is by chance correct ($I_x = I_n$) and the person can fake some DNA keys, however, the DNA keys cannot produce the $I_n$ when cross-referencing with other signature materials (figure **4A**). For example, the DNA keys from X can reveal $I_n$ only with counterfeit item-X, but not with authentic item-C (figure **4B**). Similarly, the DNA keys from C can reveal $I_n$ with item-C, but not with item-X. Therefore, the reverse decryption approach will have little impact on the safety of the DNA keys, as it cannot pass the cross-referencing test.

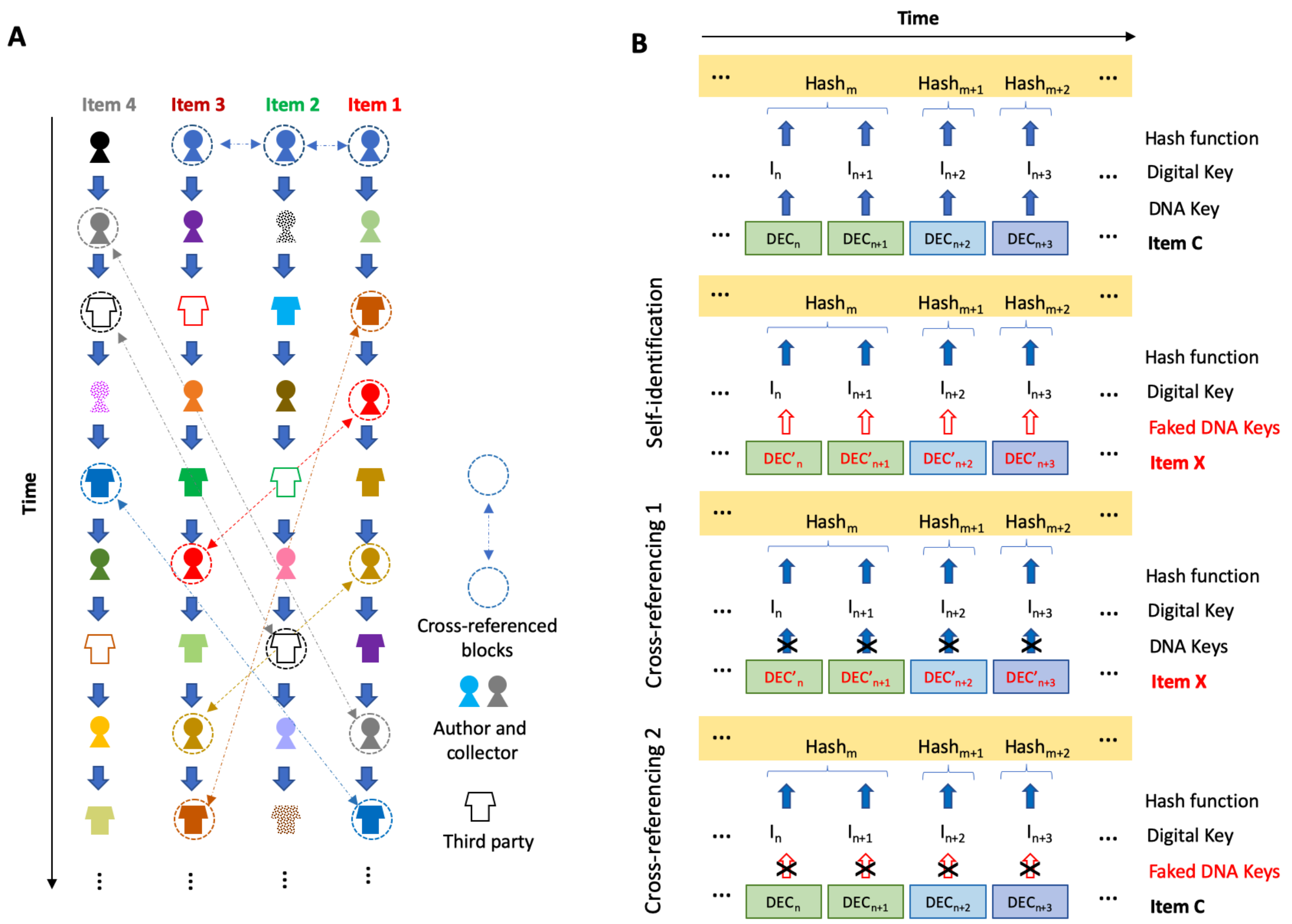

*Figure **4**. **Block cross-referencing**. **A**) When different items are associated with the same DEC, of which the hash value and association (e.g. with the author, collector of third party) had been published, the DEC can be further verified through cross-referencing. For example, the author uses the same DEC for items 1-3, and a different DEC for item 4. The author's DECs for items 1-3 can be cross-referenced with each other. **B**) The DNA keys from X can reveal $I_n$ only with counterfeit item-X, but not with authentic item-C. Similarly, the DNA keys from C can reveal $I_n$ only with item-C, but not with item-X.*

In summary, I propose the design of a signature material by combining cryptographic hash function with DNA steganography. Both methods can generate highly complex and safe keys. While the DNA-based molecular keys can be physically connected to the item to be certificated, the digital keys associated with published hash values can make the recording system transparent and facilitate cross-referencing. Both digital key and DNA keys for a block is not limited to the SHA-2 functions (e.g. SHA-256 and SHA-512) and the DNA steganography[14] used in this paper as examples. Blocks could be cross-referenced, when author, collector, or third party uses the same DEC for different items. While the block chain will grow over time, new hash function and DNA steganography will be incorporated to improve the security over time. For example, in addition to primers and restriction enzymes as DNA keys, the CRISPR/Cas-based system could provide another attractive mechanism to enhance key complexity, by adding more layers of covering DNA, which can be removed specifically by CRISPR RNA sequence-dependent Cas *trans*-cleavage activity[12]. Most importantly, the efficacy of P-D-H protocol is based on the combination of a physical object, an experimental procedure, and a digital key as a signature. Although automation has become increasingly used in biotechnology, often known as high throughput methods, their speeds lag far behind the silicon-based computations. Therefore, if a DNA steganography possesses the same bits as a digital cryptography, the DNA keys are much less vulnerable than the digital keys against brutal force search.

## Acknowledgement

I would like to thank Dr. Luca Mannocci for insightful discussions. This study is supported by BMBF (German Federal Ministry of Education and Research) grant metaZIK “OptiZeD” number 03Z22E511.

## Supporting information

# Examples of generating hash values

**$I_1$**: ACGTTGCGCAATGCCCGTAA
**$I_2$**. AATTGGCATCGGGAACTCAC
**$I_3$**: ATCTACTTCATGGGTCGACG
**$I_4$**: GTTGAACGCTACGTACATGC
**$I_5$**: TCCTAAATTCAGGGTAGTGG

**$Hash_1$**=SHA256(ACGTTGCGCAATGCCCGTAAAATTGGCATCGGGAACTCACATCTACTTCATGGGTCGACGGTTGAACGCTACGTACATGCTCCTAAATTCAGGG TAGTGG) = 68a755bf382f472b898601346ce2d75f9850526bca3e4c9e314f2956e8d12e9a

**$I_6$**: TATACCTCGGTCTGATATGC
**$I_7$**: TCTCAATCTAGGTTGTCGGT
**$I_8$**: GGGTGTAAACTCGTATGTCA

**$Hash_2$**=SHA256(68a755bf382f472b898601346ce2d75f9850526bca3e4c9e314f2956e8d12e9aTATACCTCGGTCTGATATGCTCTCAATCTAGGTTGTCG GTGGGTGTAAACTCGTATGTCA) = 2b92740fbd95714accaba38dea21fcf8cbbcffa2b1d953a65446b2f24227d1f7

**$I_9$**: CCTCTATCTATGGTCTGTCT

**$Hash_3$**=SHA256(2b92740fbd95714accaba38dea21fcf8cbbcffa2b1d953a65446b2f24227d1f7CCTCTATCTATGGTCTGTCT) = ffb0b6926d08fe9fc86a9c59362bec6fc451eec0f5c6c0f5e2dc915e3979aba5